\documentclass[twocolumn,superscriptaddress,amsmath,ammsymb]{revtex4-1}
\usepackage{epsfig}
\usepackage{amsmath}
\usepackage{amssymb}
\usepackage{graphicx}
\usepackage{dcolumn}
\usepackage{bm}
\usepackage[usenames]{color}
\usepackage[breaklinks]{hyperref}
\usepackage{soul}
\usepackage{ulem}
\hypersetup{colorlinks=true,allcolors=blue}

\begin{document}
	
\title{Quantum oscillations and nonsaturating magnetoresistivity in nodal-line semimetals} 
\author{Rui Min}
\affiliation{School of Optoelectronic Information and Physical Science, Jiangnan University, Wuxi 214122, China}
\author{Yi-Xiang Wang}
\email{wangyixiang@jiangnan.edu.cn}
\affiliation{School of Optoelectronic Information and Physical Science, Jiangnan University, Wuxi 214122, China}
\date{\today}

\begin{abstract} 
Understanding the magnetotransport behaviors in topological systems remains alluring, as a lot of intrinsic information could be extracted, \textit{e.g.}, the band structures, Berry phase, Fermi surface, carrier density, and so on.  
Motivated by the recent magnetotransport developments in nodal-line semimetal, EuGa$_4$, in this paper, we will study the magnetotransport properties of the system, focusing on the quantum oscillations and nonsaturating magnetoresistivity (MR).    
Firstly, we analyze the chemical potential and magnetoconductivity oscillations with the magnetic field and reveal that there exist two distinct oscillation frequencies, which are caused by the characteristic torus Fermi surface and can be regarded as an important experimental signature of nodal-line semimetals. 
Then we calculate the MR and find that although the MR is nonsaturating with the magnetic field in the low-energy region, the MR ratio is much smaller than that reported in the experiment. 
\end{abstract} 

\maketitle

\section{Introduction} 

Over the last decades, the theoretical and experimental studies on topological semimetals have been witnessed as one of the most active fields in condensed matter physics~\cite{B.Q.Lv, J.A.Sobota}.  Topological semimetals are three-dimensional (3D) phases of matter with gapless excitations that are protected by topology and symmetry.
Nodal-line semimetals represent a new type of topological semimetals that are beyond  Dirac and Weyl semimetals~\cite{A.A.Burkov, M.Phillips, C.Fang}. 
In nodal-line semimetals, the bands will touch on a continuous one-dimensional (1D) line in the momentum space, and the quantized Berry phase can lead to the protected surface states, which take the shape of a drumhead~\cite{I.Belopolski, G.Biana, Y.H.Chan}; while in Dirac and Weyl semimetals, the bands only touch on discrete points.  Due to the unique band structures and surface states,  nodal-line semimetals are expected to exhibit unconventional properties, such as the electromagnetic response~\cite{S.T.Ramamurthy}, the optical excitations~\cite{S.Barati, C.Wang}, as well as the correlation effect~\cite{Y.Huh, Y.Shao, Z.Wu}.

Initially, nodal-line semimetals were proposed to be realized in a multilayer heterostructure composed of alternating layers of the topological insulators and normal insulators~\cite{A.A.Burkov, M.Phillips}.  
The density functional calculations predicted that Ca$_3$P$_2$ may be a candidate to host nodal-line semimetals~\cite{Y.H.Chan, L.S.Xie}.  
Experimentally, through the angle-resolved photoemission spectroscopy and transport studies, the nodal-line semimetal state was demonstrated to exist in a few single crystal systems, such XTaSe$_2$ (X=Tl, Pb)~\cite{G.Biana, G.Bianb}, Co$_2$MnGa~\cite{I.Belopolski}, HfSiS~\cite{D.Takane}, CaAgX (X=P, As)~\cite{Y.Okamoto, E.Emmanouilidou, Y.H.Kwan, H.T.Hirose, X.B.Wang, N.Xu}, ZrSiX (X=S, Se, Te)~\cite{L.M.Schoop, M.Neupane, J.Hu} and so on.  While Dirac and Weyl semimetals are robust to generic perturbations, the stability of the nodal lines requires additional symmetries, such as the mirror reflection~\cite{I.Belopolski, L.S.Xie, G.Biana, G.Bianb, D.Takane, Y.Okamoto, E.Emmanouilidou, Y.H.Kwan, H.T.Hirose,X.B.Wang, N.Xu} and nonsymmorphic symmetry~\cite{L.M.Schoop, M.Neupane, J.Hu}. 

Recently, the square-net compound EuGa$_4$ was identified as a magnetic nodal-line semimetal~\cite{S.Lei}, in which the line nodes form a closed ring structure around the Fermi level.  
Through the magnetotransport measurements, the nonsaturating magnetoresistivity (MR) was revealed to persist to the strong magnetic field of $40$ Tesla and the MR ratio can be as giant as $R\simeq2\times10^5\%$.  
Since the MR ratio exceeds more than two orders of magnitude when compared to other topological semimetals~\cite{M.N.Ali, C.Shekhar, X.Huang, B.Fauque}, the authors suggested that it arises as a consequence of the nodal-line semimetal state~\cite{S.Lei}.  
The existing theoretical analysis of the MR mainly employed the semiclassical Boltzmann transport equation~\cite{M.X.Yang, V.Pandey}, while a full quantum mechanical study in the framework of the Landau quantization is still lacking, which motivates the present study. 

On the other hand, the magnetic field can drive the electrons to move on curved orbits with a set of discrete energies, or called the Landau levels (LLs).  
When the magnetic field increases, the LLs will cross the chemical potential successively, leading to the quantum oscillations of the related thermodynamic and transport quantities~\cite{D.Shoenberg, L.Zhang, J.Knolle, Y.X.Wang2023}.  In nodal-line semimetals, the previous quantum oscillation studies  mainly focused on the accumulated Berry phase analysis~\cite{C.Li, L.Oroszlany, H.Yang}.  
As the phase analysis is closely related to the judgment of the intercept with the coordinate axis, it may become obscured in the experiment~\cite{Y.Zhao}.  
Thus, an interesting question arises whether unambiguous signatures of nodal-line semimetals can be extracted from the quantum oscillations?  
 
In this paper, we will adopt a simple torus model to describe the nodal-line semimetal state.  Firstly, we analyze the band structures and LLs under a magnetic field.  
Then under the condition of fixed carrier density, we study the chemical potential evolution with the magnetic field, and further based on the Kubo-Bastin formula, we calculate the magnetoconductivity (MC) and MR.  The results in the low-energy region and high-energy region are compared.  
Meanwhile, the effects of the impurity scatterings on the magnetotransport are also taken into account.    
The main results are given as follows: (i) In the chemical potential and MC oscillations with the inverse magnetic field, there exist two distinct oscillation frequencies in the low-energy region, but only one in the high-energy region.  
The two distinct oscillation frequencies are attributed to the characteristic torus Fermi surface, and can be regarded as an important signature of nodal-line semimetals.  
This conclusion is supported by several experimental studies on the de Haas-van Alphen (dHvA) oscillations of nodal-line semimetals~\cite{J.Hu, Y.H.Kwan}.  
(ii) The calculations indicate that the nonsaturating MR with the magnetic field occurs only in the low-energy region, but not in the high-energy region. 
Furthermore, the MR ratio $R$ lies within the range $(50\%\sim400\%)$ over a large variation of the Dirac mass and the carrier density.  However, the ratio is much smaller than that reported in the experiment~\cite{S.Lei}.  
Our work could provide more deep insights into the magnetotransport behaviors observed in nodal-line semimetal experiments.

\section{Model}

\begin{figure}
	\includegraphics[width=9cm]{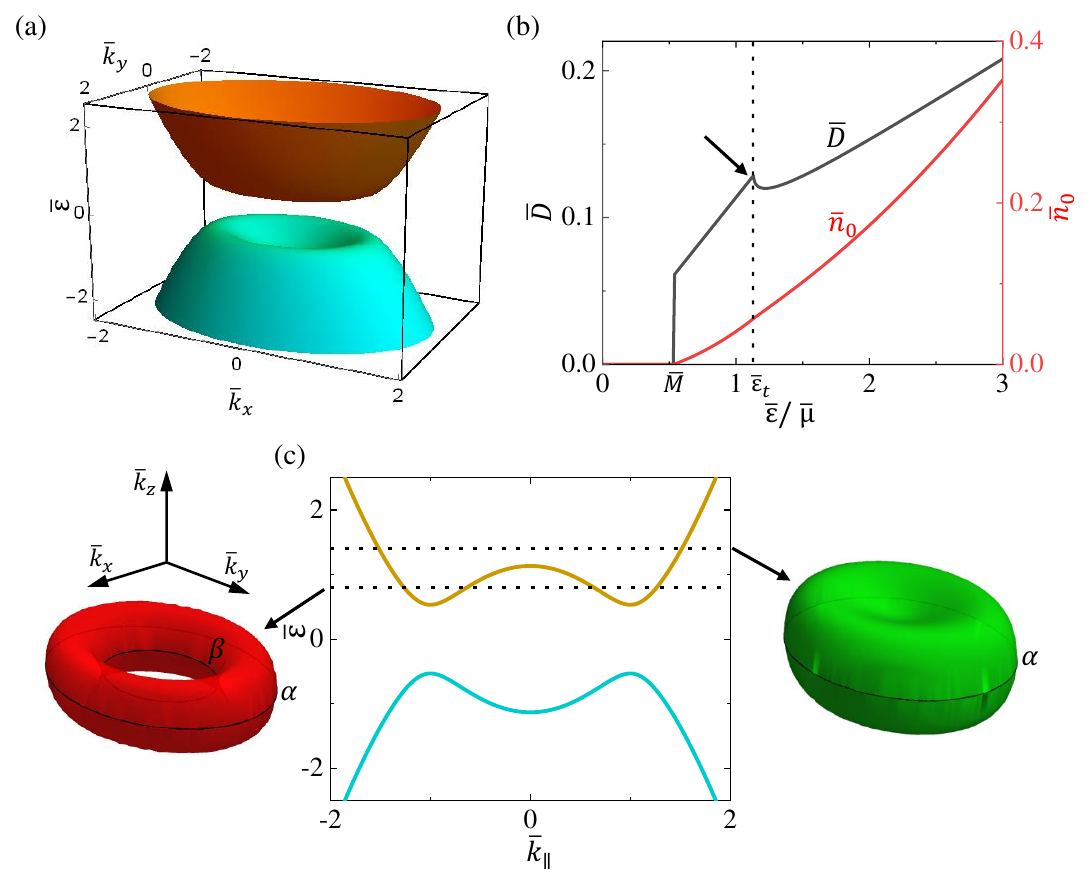}
	\caption{(Color online)  (a) The 2D bands of nodal-line semimetals in the $(\bar k_x,\bar k_y)$ space with $\bar k_z=0$.  (b) The DOS $\bar D(\bar\varepsilon)$ and carrier density $\bar n_0(\bar\mu)$.  Note the double $y$-axis in the figure.  The DOS shows a singularity at the transition point   $\bar\varepsilon=\bar\varepsilon_t$, as indicated by the arrow.  
	(c) The simplified 1D bands vs $\bar k_\parallel$, with $\bar k_z=0$.  The left and right insets show the 3D Fermi surface in the momentum space, in which the Fermi energy $\bar\varepsilon=0.8$ and $\bar\varepsilon=1.4$, respectively.  Note that the extremal circular orbits, $\alpha$ and $\beta$, are denoted by the  black lines. }  
	\label{Fig1}
\end{figure}

Consider a simple torus model to describe the nodal-line semimetal state~\cite{S.Barati, Y.H.Chan, C.Wang, Y.Huh}, with the Hamiltonian written as 
\begin{align}
\hat H(\boldsymbol k)=\Big(\frac{\hbar^2 k_\parallel^2}{2m}-\varepsilon_r\Big)\sigma_x+\hbar v_zk_z\sigma_y+M\sigma_z,  
\end{align}
where $\boldsymbol \sigma$ are the Pauli matrices acting on the pseudospin space. 
$k_\parallel=\sqrt{k_x^2+k_y^2}$ is the radial wave vector in the $x$-$y$ plane, $m$ is the effective electron mass, $\varepsilon_r=\frac{\hbar^2 k_r^2}{2m}$ controls the size of the nodal ring, with $k_r$ being the radius, $v_z$ is the Fermi velocity in the $z$ direction, and $M$ is the Dirac mass.  
When the Dirac mass is absent, $M=0$, the nodal lines are protected by the combined inversion and time-reversal symmetry, in which the operators are given as $\hat{\cal P}=\sigma_x$ and $\hat{\cal T}=\hat{\cal K}$, with $\hat{\cal K}$ being the complex conjugate operator, exhibiting the parity anomaly~\cite{C.Wang, W.B.Rui}.  The introduction of the Dirac mass term $M\sigma_z$ will break $\hat{\cal P}\hat{\cal T}$, leading to the gap opening of the bands.  

The energies and wavefunctions of $\hat H(\boldsymbol k)$ are solved as    
\begin{align}
&\bar\varepsilon_s(\bar{\boldsymbol k})
=s\sqrt{(\bar k_\parallel^2-1)^2+\gamma^2\bar k_z^2+\bar M^2},
\end{align}
and
\begin{align}
&u_+(\bar{\boldsymbol k})=\begin{pmatrix}
\text{cos}\frac{\theta_{\bar k}}{2}\\ 
\text{sin}\frac{\theta_{\bar k}}{2}e^{i\varphi_{\bar k}}
\end{pmatrix}, \quad 
u_-(\bar{\boldsymbol k})=\begin{pmatrix}
\text{sin}\frac{\theta_{\bar k}}{2}\\ -\text{cos}\frac{\theta_{\bar k}}{2}e^{i\varphi_{\bar k}}
\end{pmatrix}, 
\end{align}
respectively.  Here the index $s =\pm1$ denotes the conduction/valence band, the angles $\theta_{\bar k}=$tan$^{-1} \frac{\sqrt{(\bar k_\parallel^2-1)^2+\gamma^2\bar k_z^2}}{\bar M}$ and $\varphi_{\bar k}=$tan$^{-1}\frac{\gamma\bar k_z}{\bar k_\parallel^2-1}$. 
To simplify the calculations below, we will use $k_r$ and $\varepsilon_r$ as the scale of the wave vector and energy, respectively, and define the dimensionless quantities as $\bar k_\parallel=\frac{k_\parallel}{k_r}$, $\bar k_z=\frac{k_z}{k_r}$, $\bar M=\frac{M}{\varepsilon_r}$, $\bar\varepsilon=\frac{\varepsilon}{\varepsilon_r}$, and $\gamma=\frac{2mv_z}{\hbar k_r}$. 
In Fig. 1(a), the 2D bands are plotted in the $(\bar k_x,\bar k_y)$ space with $\bar k_z=0$, where a pair of nodal rings with unit radius $\bar k_r=1$ exist in the conduction and valence bands.  In our calculations, unless specified, the model parameters are set as $\bar M=0.53$ and $\gamma= 0.7$.  

For the $s$-band, the Berry curvature is defined as $\Omega_s(\bar{\boldsymbol k})=\nabla_{\bar{\boldsymbol k}}\times A_s(\bar{\boldsymbol k})$, where the Berry connection $A_s(\bar{\boldsymbol k})=\langle u_s(\bar{\boldsymbol k})|i\nabla_{\bar{\boldsymbol k}}|u_s(\bar{\boldsymbol k})\rangle$.  After a direct calculation, the Berry curvature is given as~\cite{W.B.Rui}
\begin{align}
\Omega_s(\bar{\boldsymbol k})=\frac{s\gamma\bar k_\parallel\bar M}{\big[(\bar k_\parallel^2-1)^2+\gamma^2\bar k_z^2+\bar M^2\big]^{3/2}}\hat{\boldsymbol e}_\phi.  
\end{align}
Clearly, when the Dirac mass $\bar M=0$, $\Omega_s$ vanishes in the whole BZ; but
on the Dirac nodal line, $\Omega_s(\bar{\boldsymbol k})=s\pi\delta(\bar k_\parallel-\bar k_r)\delta(\bar k_z)\hat{\boldsymbol e}_\phi$ and becomes singular.  A finite $\bar M$ will regularize the Berry curvature.  
The Berry phase is obtained by integrating over the Berry curvature 
$I_s=\int d\bar k_\parallel d\bar k_z \Omega_s(\bar{\boldsymbol k})=s\pi$, indicating the 
topologically nontrivial bands.  Such nontrivial bands in nodal-line semimetals can be reflected in the Drude weight~\cite{S.Barati} as well the Hall coefficient with no magnetic field~\cite{W.B.Rui}.

In a 3D system, with the help of the Green's function $\hat G(\boldsymbol k,z)=[z-\hat H(\boldsymbol k)]^{-1}$, the DOS is defined as
\begin{align}
D(\varepsilon)=-\frac{1}{\pi V}\sum_{\boldsymbol k}
\text{Im}\hat G(\boldsymbol k,z=\varepsilon+i0^+), 
\label{DOS}
\end{align}
where $V$ is the volume.  In nodal-line semimetals, the dimensionless DOS $\bar D(\bar\varepsilon)=\frac{D(\varepsilon)}{k_r^3/\varepsilon_r}$ can be obtained analytically,  
\begin{align} 
\bar D(\bar\varepsilon>0)=\frac{\bar\varepsilon}{4\pi\gamma}\Big[
\Theta(\bar\varepsilon-\bar M)-\frac{1}{\pi}\text{tan}^{-1}
\sqrt{\bar\varepsilon^2-\bar\varepsilon_t^2}\Theta\big(\bar\varepsilon-\bar\varepsilon_t\big)
\Big],  
\end{align}
where $\Theta(x)$ is the step function and $\bar\varepsilon_t=\sqrt{\bar M^2+1}$ denotes the energy at the  transition point that separates the low-energy region $|\bar\varepsilon|<\bar\varepsilon_t$ and high-energy region $|\bar\varepsilon|>\bar\varepsilon_t$.  

In Fig.~\ref{Fig1}(b), we plot the DOS $\bar D$ as a function of the energy $\bar\varepsilon$.  
The DOS exhibits a nonmonotonic variation that can be understood from the Fermi surface evolutions.  
In the low-energy region, the torus Fermi surface can have two extremal circular orbits: the outer $\alpha$ orbit and the inner $\beta$ orbit [Fig.~\ref{Fig1}(c), the left inset], with  
the radius given as $\bar k_\alpha=\big[(\bar\varepsilon^2-\bar M^2)^\frac{1}{2}+1\big]^\frac{1}{2}$ and $\bar k_\beta=\big[-(\bar\varepsilon^2-\bar M^2)^\frac{1}{2}+1\big]^\frac{1}{2}$, respectively.    
With increasing $\bar\varepsilon$, the $\alpha$ orbit enlarges and the $\beta$ orbit shrinks, leading to the linear dependence of the DOS on $\bar\varepsilon$.  
At the transition point $\bar\varepsilon=\bar\varepsilon_t$, we have $\bar k_\beta=0$, meaning the occurrence of the Lifshitz transition of the Fermi surface.  Correspondingly, the DOS shows an abrupt change, as indicated by the arrow in Fig.~\ref{Fig1}(b).  When increasing $\bar\varepsilon$ to enter the high-energy region, the Fermi surface includes only the $\alpha$ orbit [Fig.~\ref{Fig1}(c), the right inset], and $\bar D(\bar\varepsilon)$ further increases.  

\begin{figure}
	\includegraphics[width=9.2cm]{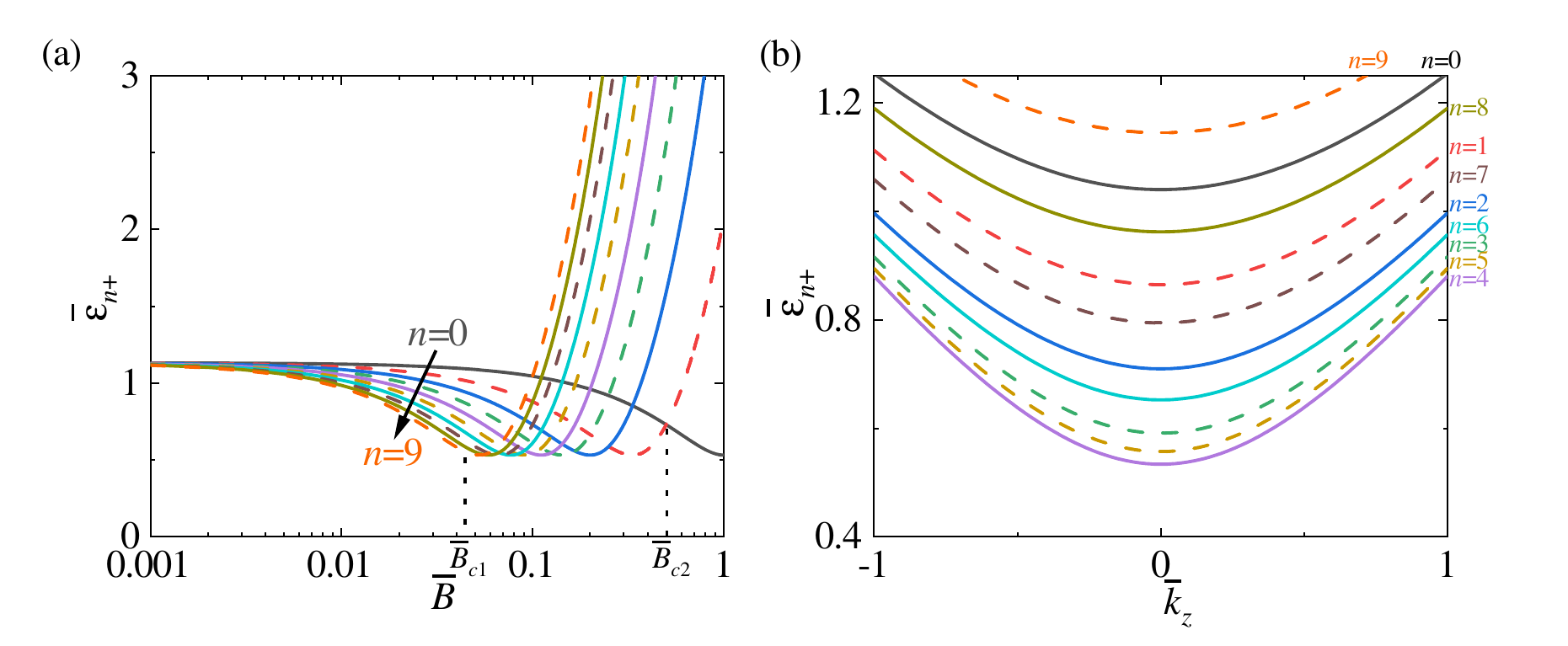}
	\caption{(Color online) (a) The LL spectra vs the magnetic field $\bar B$. The arrow indicates the LL index growing from $n=0$ up to $n=9$.  The dotted lines label the critical fields $\bar B=\bar B_{c1}$ and $\bar B=\bar B_{c2}$.  (b) The LL dispersion vs $\bar k_z$ when $\bar B=0.106$, with the LL index $n$ being labeled.  In (a) and (b), each LL owns a definite even or odd parity that are represented by the solid or dashed lines, respectively. }  
	\label{Fig2}
\end{figure} 

For the carrier density $n_0$, it is related to the DOS as 
\begin{align}
n_0(\mu)=\int_0^\infty d\varepsilon D(\varepsilon)f(\varepsilon)
+\int_{-\infty}^0 d\varepsilon D(\varepsilon)[f(\varepsilon)-1],  
\label{n0}
\end{align}
where $f(\varepsilon)=\frac{1}{\text{exp}[\beta(\varepsilon-\mu)]+1}$ is the Fermi distribution function, $\beta=\frac{1}{k_BT}$ is the inverse temperature, and $\mu$ denotes the chemical potential.  At zero temperature $T=0$, the dimensionless carrier density $\bar n_0=\frac{n_0}{k_r^3}$ is obtained as 
\begin{align}
\bar n_0(\bar\mu)=&\frac{1}{8\pi\gamma}\Big[
(\bar\mu^2-\bar M^2)\Theta(\bar\mu-\bar M) +\Big(\frac{1}{\pi}\sqrt{\bar\mu^2-\bar\varepsilon_t^2}
\nonumber\\
&-\frac{1}{\pi}(\bar\mu^2-\bar M^2)\text{tan}^{-1} \sqrt{\bar\mu^2-\bar\varepsilon_t^2}\Big)  
\Theta\big(\bar\mu-\bar\varepsilon_t\big)\Big],  
\label{n0mu}  
\end{align}
where the dimensionless chemical potential is defined as $\bar\mu=\frac{\mu}{\varepsilon_r}$.  In Fig.~\ref{Fig1}(b), the carrier density $\bar n_0$ is also plotted as a function of $\bar\mu$. 
We see that $\bar n_0$ exhibits the nonlinear increasing with $\bar\mu$, and behaves quite smooth when $\mu$ crosses the transition point. 
For the chosen $\bar\mu=0.8$ and $\bar\mu=1.4$, Eq.~(\ref{n0mu}) indicates that the corresponding carrier density is $\bar n_0=0.02042$ and $\bar n_0=0.08946$, respectively.

\section{Landau Levels}

When a perpendicular magnetic field $\boldsymbol B=B\boldsymbol e_z$ is applied on the system, the quantized LLs will be formed and disperse along the $z$ direction.  
To solve the LLs, we include the magnetic field in the system through the Peierls substitution, $\boldsymbol k\rightarrow\boldsymbol\pi=\boldsymbol k+e\boldsymbol A$. 
In the Landau gauge, the magnetic vector potential $\boldsymbol A$ is chosen as $\boldsymbol A=(-By,0,0)$.  
After introducing the ladder operators $\hat a=\frac{l_B}{\sqrt2}(\pi_x-i\pi_y)$ and $\hat a^\dagger=\frac{l_B}{\sqrt2}(\pi_x+i\pi_y)$, with the magnetic length $l_B=\sqrt{\frac{\hbar}{eB}}$,  the Hamiltonian becomes 
\begin{align}
\hat{\bar H}_B=\begin{pmatrix}
\bar M& \frac{2}{l_B^2} \hat a^\dagger a+\frac{1}{l_B^2}-1-i\gamma\bar k_z
\\
\frac{2}{l_B^2} \hat a^\dagger a+\frac{1}{l_B^2}-1+i\gamma\bar k_z& -\bar M   
\end{pmatrix}. 
\end{align}

By using the trial wavefunction $\psi_n=\begin{pmatrix}
c_n^1|n\rangle\\ c_n^2|n\rangle
\end{pmatrix}$, where the states $|n\rangle$ are the eigenstates of the harmonic oscillator, $\hat a^\dagger\hat a|n\rangle=n|n\rangle$, and $c_n^{1,2}$ denote the coefficients, the energies and eigenfunctions for the $n$-th LLs are solved as 
\begin{align}
&\bar\varepsilon_{ns}
=s\sqrt{\big(2n\bar B+\bar B-1\big)^2+\bar M^2+\gamma^2\bar k_z^2}, 
\end{align}
and
\begin{align}
&\psi_{n+}=\begin{pmatrix}
\text{cos}\frac{\theta_n}{2} |n\rangle\\
\text{sin}\frac{\theta_n}{2} e^{i\varphi_n}|n\rangle
\end{pmatrix}, \quad 
\psi_{n-}=\begin{pmatrix}
\text{sin}\frac{\theta_n}{2} |n\rangle\\
-\text{cos}\frac{\theta_n}{2} e^{i\varphi_n}|n\rangle
\end{pmatrix},   
\end{align}
respectively.  Here the angles $\theta_n=\text{tan}^{-1}\frac{\sqrt{(2n\bar B+\bar B-1)^2+\gamma^2\bar k_z^2}}{\bar M}$, $\varphi_n=\text{tan}^{-1}\frac{\gamma\bar k_z}{2n\bar B+\bar B-1}$, and the dimensionless magnetic field is defined as $\bar B=\frac{B}{l_{B0}^2 k_r^2\cdot 1\text{ T}}$, in which $l_{B0}=25.6$ nm denotes the magnetic length at the unit magnetic field $B=1$ T.

In Fig.~\ref{Fig2}(a), we display the LL spectra as a function of the magnetic field $\bar B$, with its index growing from $n=0$ up to $n=9$.  
As the Hamiltonian $\bar H_B$ owns the parity symmetry with the operator $\hat{\cal P}=(-1)^{\hat a^\dagger\hat a}I$, each LL carries a definite even or odd parity~\cite{T.Devakul, Y.X.Wang2022} that is represented by the solid or dashed line.  
We see that with increasing $\bar B$, each LL decreases at first and then increases.  
Such a nonmonotonic variation will change the LL sequence.  As shown in Fig.~\ref{Fig2}(a), when $\bar B<\bar B_{c1}=0.045$, the $n=0$ LL is located at the top and the $n=9$ LL at the bottom; when $\bar B_{c1}<\bar B<\bar B_{c2}$, the $n\geq1$ LLs will cross its minimum at $\bar B=\frac{1}{2n+1}$ successively, leading to a chaotic sequence of the LLs, as seen in Fig.~\ref{Fig2}(b) with $\bar B=0.106$; when $\bar B>\bar B_{c2}=0.5$, the positions of the LLs from $n=0$ to $n=9$ will be reversed, in which the $n=0$ LL will be located at the bottom and the $n=9$ LL at the top.

\section{Chemical Potential} 

\begin{figure}
	\includegraphics[width=8.2cm]{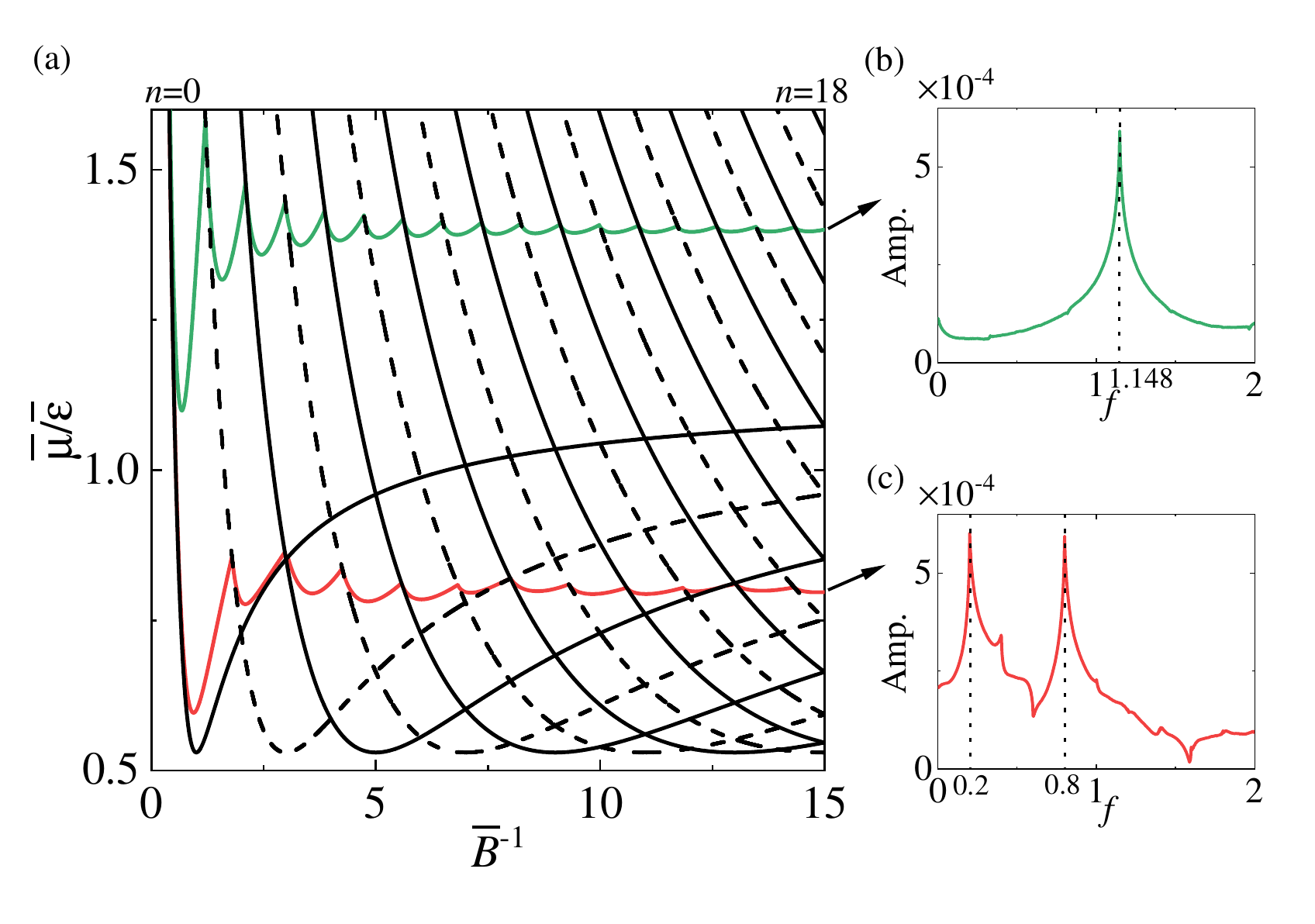}
	\caption{(Color online) (a) The evolution of the chemical potential $\bar\mu$ with the inverse magnetic field $\bar B^{-1}$.  The red and green lines are under fixed carrier densities 
	$\bar n_0=0.02042$ and $0.08946$, respectively.  
	The LL spectra are also plotted, with the index from $n=0$ up to $n=18$, in which the even and odd parities are represented by the solid and dashed lines, respectively.  (b)-(c) The FFT analysis of the chemical potential.} 
	\label{Fig3}
\end{figure}

Next, we study the chemical potential oscillations with the magnetic field in nodal-line semimetals.  According to Eq.~(\ref{DOS}), the DOS of the system under a magnetic field is given as 
\begin{align}
\bar D(\bar\varepsilon,\bar B)=\frac{\bar B}{2\pi^2\gamma}\sum_{n\geq 0} 
\frac{\bar\varepsilon}{\sqrt{\bar\varepsilon^2-\big(2n\bar B+\bar B-1\big)^2-\bar M^2}}.  
\end{align} 
Note that the summations over $k_x$ and $k_y$ give the degeneracy factor of each LL,  $\sum_{k_x,k_y}=\frac{L_xL_y}{2\pi l_B^2}=\frac{\bar L_x\bar L_y\bar B}{2\pi}$, with the dimensionless length $\bar L_{x(y)}=L_{x(y)} k_r$. 
From Eq.~(\ref{n0}), at $T=0$, the carrier density is obtained as 
\begin{align}
\bar n_0(\bar\mu,\bar B)=\frac{\bar B}{2\pi^2\gamma}
\sum_{n\geq0}\sqrt{\bar \mu^2-\big(2n\bar B+\bar B-1\big)^2-\bar M^2}. 
\label{n0B}
\end{align}  
Under fixed carrier density $\bar n_0$, the chemical potential $\bar\mu$ can be obtained through solving Eq.~(\ref{n0B}).  In Fig.~\ref{Fig3}(a), we plot the results of $\bar\mu$ as a function of the inverse magnetic field $\bar B^{-1}$.  

In the high-energy region with fixed carrier density $\bar n_0=0.08946$, the chemical potential $\bar\mu$ exhibits evident oscillations with $\bar B^{-1}$ around $\bar\mu=1.4$ [Fig.~\ref{Fig3}(a), the green line].  
The oscillation peaks in $\bar\mu$ are due to the magnetic field-driven LL crossings over $\bar\mu$.  
To reveal the oscillation frequency, we perform the FFT analysis.  
As seen in Fig.~\ref{Fig3}(b), the FFT amplitude shows one strong peak at the frequency $f=1.148$, from which the oscillation period is obtained as $\Delta_{1/\bar B}=f^{-1}=0.871$. 
On the other hand, according to the Onsager's relation, the oscillation period $\Delta_{1/\bar B}^O$ is inversely proportional to the extremal Fermi surface area $\bar S_{\text{e}}$ (expressed in the dimensionless quantities)~\cite{D.Shoenberg},  
\begin{align}  
\Delta_{1/\bar B}^O=\frac{2\pi e}{\hbar\bar S_{\text{e}}} l_{B0}^2 \cdot 1\text{ T}.  
\label{Onsager}
\end{align}
As the area spanned by the $\alpha$ orbit is $\bar S_{\text{e}\alpha}=7.212$ [Fig.~\ref{Fig1}(c), the right inset], the period is determined as $\Delta_{1/\bar B}^{O\alpha}=0.871$, which matches the FFT result $\Delta_{1/\bar B}$.  

In the low-energy region with fixed $\bar n_0=0.02042$, $\bar\mu$ oscillates around $\bar\mu=0.8$ [Fig.~\ref{Fig3}(a), the red line].  In Fig.~\ref{Fig3}(c) of the FFT analysis, it is interesting to find that the FFT amplitude exhibits two strong peaks at the frequencies $f_a=0.8$ and $f_b=0.2$, from which two periods are obtained as $\Delta_{1/\bar B}^a=f_a^{-1}=1.25$ and $\Delta_{1/\bar B}^b=f_b^{-1}=5$. 
The existence of two oscillation frequencies are caused by the characteristic torus Fermi surface in nodal-line semimetals.  
Now since the torus Fermi surface owns two extremal $\alpha$ and $\beta$ circular orbits, the experimental responses will be a superposition of the two oscillations.  
The areas spanned by the $\alpha$ and $\beta$ orbits are given as $\bar S_{\text{e}\alpha}=5.024$ and $\bar S_{\text{e}\beta}=1.256$, respectively [Fig.~\ref{Fig1}(c), the left inset].  
Then by using the Onsager's relation in Eq.~(\ref{Onsager}), two periods are determined as $\Delta_{1/\bar B}^{O\alpha}=1.25$ and $\Delta_{1/\bar B}^{O\beta}=5$, which also match the FFT result.  

Intuitively, when $\bar B^{-1}$ increases, each LL will go through the process of first decreasing and then increasing.  
In the low-energy region, both the LL decreasing and increasing processes can cross over $\bar\mu$.  As each process owns a frequency, two distinct oscillation frequencies will be found in the chemical potential.  
For comparison, in the high-energy region, the LL decreasing process can fall below $\bar\mu$, while the LL increasing process cannot cross over $\bar\mu$, thus only one oscillation frequency is left.   
The characteristic frequencies also exist in the Shubnikov-de Haas (SdH) oscillations of the MCs, as will be studied in the next section.

\section{Magnetoconductivity}

The magnetotransport measurements can provide a powerful tool to reveal the intrinsic properties of topological systems.  In this section, we study the MCs in nodal-line semimetals.  
In a 3D system, the conductivity tensors $\sigma_{\alpha\beta}$ can be calculated by using the Kubo-Bastin formula~\cite{A.Bastin, L.Smrcka}, 
\begin{align}
\sigma_{\alpha\beta}=&\frac{\hbar}{2\pi V} \sum_{\boldsymbol k}
\int_{-\infty}^\infty d\varepsilon f(\varepsilon)\Big[
\text{Tr}\big(\hat J_\alpha\frac{d\hat G^R}{d\varepsilon}J_\beta(\hat G^A-\hat G^R)
\nonumber\\
&-\hat J_\alpha(\hat G^A-\hat G^R)\hat J_\beta\frac{d\hat G^A}{d\varepsilon}\big)\Big],    
\label{Kubo-Streda}
\end{align}
in which $\hat J_\alpha=\frac{e}{\hbar}\frac{\partial\hat H}{\partial k_\alpha}$ is the current density operator, $\hat G^{R/A}(\boldsymbol k,\varepsilon)=[\varepsilon-\hat H(\boldsymbol k)\pm i\eta]^{-1}$ is the retarded/advanced Green's function, $\eta=\frac{\hbar}{\tau}$ represents the linewidth broadening induced by the impurity scatterings and $\tau$ is the scattering rate.  

\begin{figure}
	\includegraphics[width=9.2cm]{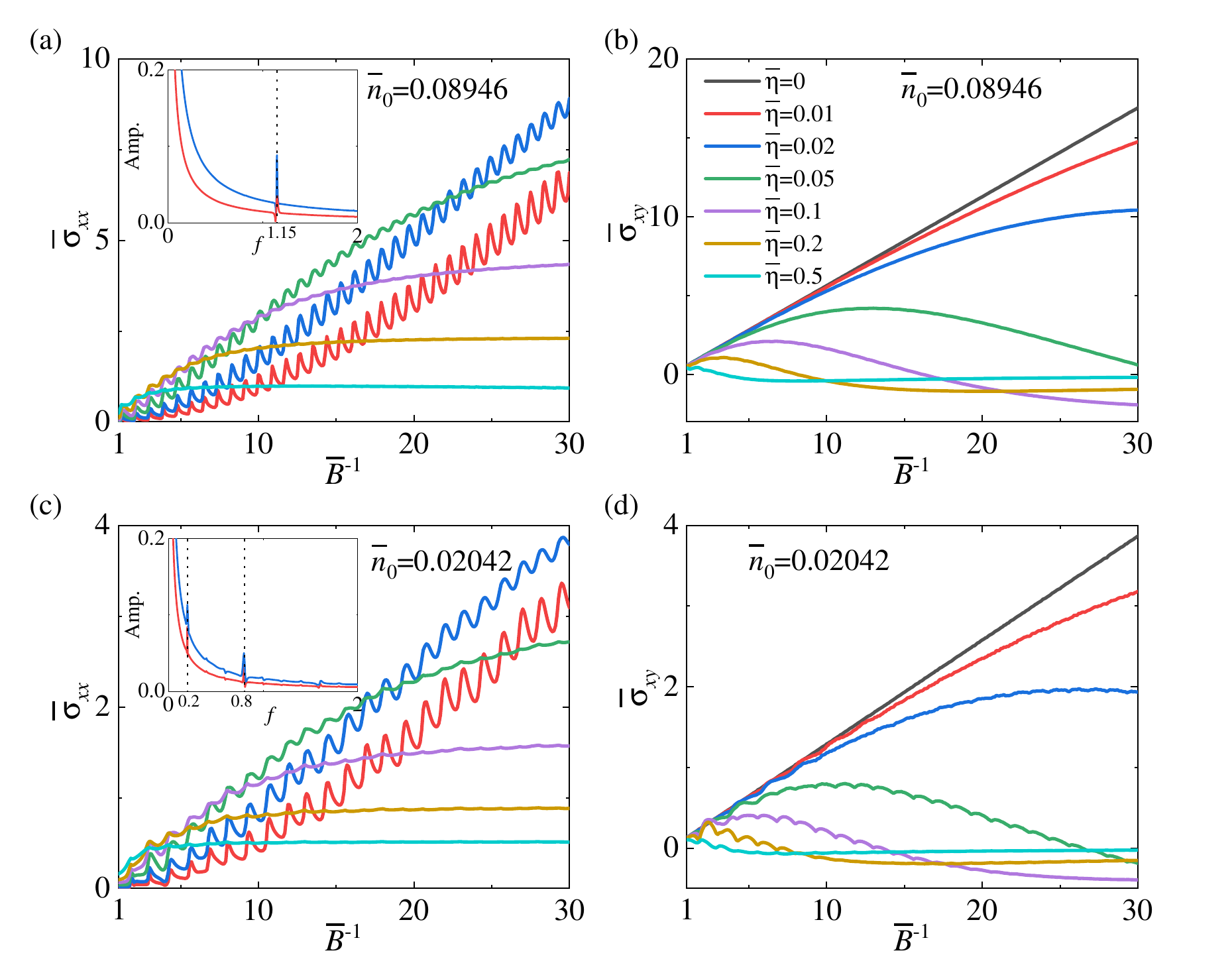}
	\caption{(Color online) The longitudinal conductivity $\bar\sigma_{xx}$ in (a) and (c), and the Hall conductivity $\bar\sigma_{xy}$ in (b) and (d) vs the inverse magnetic field $\bar B^{-1}$.  
	The different lines represent the different linewidths $\bar\eta$.  Note that $\bar\eta$ can take zero in (b) and (d), but not in (a) and (c).  
	The carrier density is fixed as $\bar n_0=0.08946$ in (a)-(b), and $\bar n_0=0.02042$ in (c)-(d).  In (a) and (c) insets, we show the FFT analysis of $\bar\sigma_{xx}$ at $\bar\eta=0.01$ and $\bar\eta=0.02$.  The legends are the same in all figures. }
	\label{Fig4}
\end{figure}

For nodal-line semimetals under a magnetic field, after inserting the LL eigenvectors $|ns\rangle$ in Eq.~(\ref{Kubo-Streda}) and making some straightforward derivations, the dimensionless longitudinal conductivity $\bar\sigma_{xx}=\frac{\sigma_{xx}}{\sigma_0 k_r}$ and Hall conductivity  $\bar\sigma_{xy}=\frac{\sigma_{xy}}{\sigma_0 k_r}$ are obtained as~\cite{Y.X.Wang2023, Z.Cai}
\begin{align} 
&\bar\sigma_{xx}=\frac{4\bar B^2\bar\eta^2}{\pi\bar L_z} 
\sum'\frac{|{\cal M}_{ns;n+1,s'}|^2}{[(\bar\mu-\bar\varepsilon_{ns})^2+\bar\eta^2]
[(\bar\mu-\bar\varepsilon_{n+1,s'})^2+\bar\eta^2]}, 
\label{sigmaxx}
\end{align} 
and
\begin{align}
&\bar\sigma_{xy}=\frac{4\bar B^2}{\bar L_z} 
\sum'\frac{(\bar\varepsilon_{ns}-\bar\varepsilon_{n+1,s'})^2-\bar\eta^2}
{[(\bar\varepsilon_{ns}-\bar\varepsilon_{n+1,s'})^2+\bar\eta^2]^2} |{\cal M}_{ns;n+1,s'}|^2
\nonumber\\
&\times[\theta(\bar\mu-\bar\varepsilon_{ns})\theta(\bar\varepsilon_{n+1,s'}-\bar\mu)
-\theta(\bar\mu-\bar\varepsilon_{n+1,s'})\theta(\bar\varepsilon_{ns}-\bar\mu)],   
\label{sigmaxy}
\end{align}
respectively.  Here $\bar L_z=L_zk_r$, $\sigma_0=\frac{e^2}{h}=(25.8\text{ k}\Omega)^{-1}$ is the unit of the quantum conductivity.  
The abbreviated summation sign is 
\begin{align}
\sum'=\sum_{\bar k_z}\sum_{n\geq 0}\sum_{s,s'}, 
\end{align}
and the matrix element is 
\begin{align}
{\cal M}_{ns;n+1,s'}=\sqrt{n+1}(c_{ns}^{2*}c_{n+1,s'}^1+c_{ns}^{1*}c_{n+1,s'}^2).  
\end{align}
Note that the nonvanishing matrix element $\langle ns|\hat J_{x/y}|n's'\rangle$ determines the common selection rules $n\rightarrow n\pm1$ in 2D and 3D Dirac systems.  

In Fig.~\ref{Fig4}, the numerical results of $\bar\sigma_{xx}$ and $\bar\sigma_{xy}$ are plotted as functions of the inverse magnetic field $\bar B^{-1}$. 
In the high-energy region, for $\bar\sigma_{xx}$ in Fig.~\ref{Fig4}(a), the SdH oscillations are clearly observed at weak $\bar\eta$.  In Fig.~\ref{Fig4}(a) inset, we display the FFT results of $\bar\sigma_{xx}$ at $\bar\eta=0.01$ and $\bar\eta=0.02$.  
We see that the FFT amplitude lies on a gradually decreasing background and exhibits one strong peak at the frequency $f=1.15$, which is the same as that in the chemical potential oscillations.  When $\bar\eta$ increases, the impurity scatterings will smoothen the SdH oscillations and drive $\bar\sigma_{xx}$ to reach its saturation value at smaller $\bar B^{-1}$ (larger $\bar B$), indicating that the system enters into the diffusive regime.  

For $\bar\sigma_{xy}$ in Fig.~\ref{Fig4}(b), when $\bar\eta=0$ of the clean case, a linear dependence between $\bar\sigma_{xy}$ and $B^{-1}$ is found as $\bar\sigma_{xy}=k\bar B^{-1}$, in which the coefficient is fitted as $k=0.5628$.  
By using the classical linear relationship (expressed in the dimensionless quantities)~\cite{Abrikosov, V.Konye}
\begin{align} 
\bar\sigma_{xy}=\frac{\bar n_0 e}{\bar B \sigma_0 l_{B0}^2 \cdot 1\text{ T}}, 
\end{align}
the carrier density is obtained as $\bar n_0=k\sigma_0 l_{B0}^2\cdot 1\text{ T}/e =0.0891$. 
The value agrees well with the chosen $\bar n_0$ value, which demonstrates the reliability of our numerical calculations.  With increasing $\bar\eta$, $\bar\sigma_{xy}$ will deviate from the linear dependence, get suppressed and even become negative.  
The negative $\bar\sigma_{xy}$ appears when $\bar\eta$ is larger than the energy difference between the neighboring LLs, as seen in Eq.~(\ref{sigmaxy}).  
At strong $\bar\eta>0.2$, $\bar\sigma_{xy}$ tends to be vanishing. 
In addition, we observe that the magnitude of $\bar\sigma_{xy}$ is comparable to $\bar\sigma_{xx}$. 

For comparison, in the low-energy region, the conductivities $\bar\sigma_{xx}$ and $\bar\sigma_{xy}$ in Figs.~\ref{Fig4}(c) and~\ref{Fig4}(d) exhibit similar dependences on $\bar B^{-1}$ and $\bar\eta$ as those in the high-energy region. 
But in Fig.~\ref{Fig4}(c) inset, the FFT amplitude exhibits two strong peaks at $f_a=0.2$ and $f_b=0.8$, which also agree with those in the chemical potential oscillations.  
The two distinct oscillation frequencies in the low-energy region are suggested as an important signature of nodal-line semimetals.  Note that in Fig.~\ref{Fig4}(d), there exist weak oscillations in $\bar\sigma_{xy}$ at $\bar\eta=0.05\sim0.2$, which are due to the combined effects of the magnetic field, the chemical potential and the impurity scatterings.  The weak oscillations in  $\bar\sigma_{xy}$ can also be found in a previous study on the anomalous Hall effect  of 3D pentatellurides~\cite{H.W.Wang}. 

Experimentally, the two distinct oscillation frequencies were clearly observed in the dHvA oscillations of nodal-line semimetals, CaAgAs~\cite{Y.H.Kwan} and ZrSiTe~\cite{J.Hu}, but their significance was not clearly recognized.  
In addition, the observed single oscillation frequency of another nodal-line semimetal,  ZrSiSe~\cite{J.Hu}, may be attributed to the high carrier density of the crystal sample, in which the chemical potential is located in the high-energy region.

\section{Magnetoresistivity}

Since the MR can be measured directly in the electric transport experiments, it establishes a bridge to connect the theoretical results to the experiments.  
In this section, we study the MR $\rho_{xx}$ in nodal-line semimetals, which is obtained as the tensor inversions of the conductivities.  The dimensionless MR $\bar\rho_{xx}=\rho_{xx}\sigma_0k_r$ is 
\begin{align}
\bar\rho_{xx}=\frac{\bar\sigma_{xx}}{\bar\sigma_{xx}^2+\bar\sigma_{xy}^2}.
\label{rhoxx}  
\end{align}  

\begin{figure}
	\includegraphics[width=9.2cm]{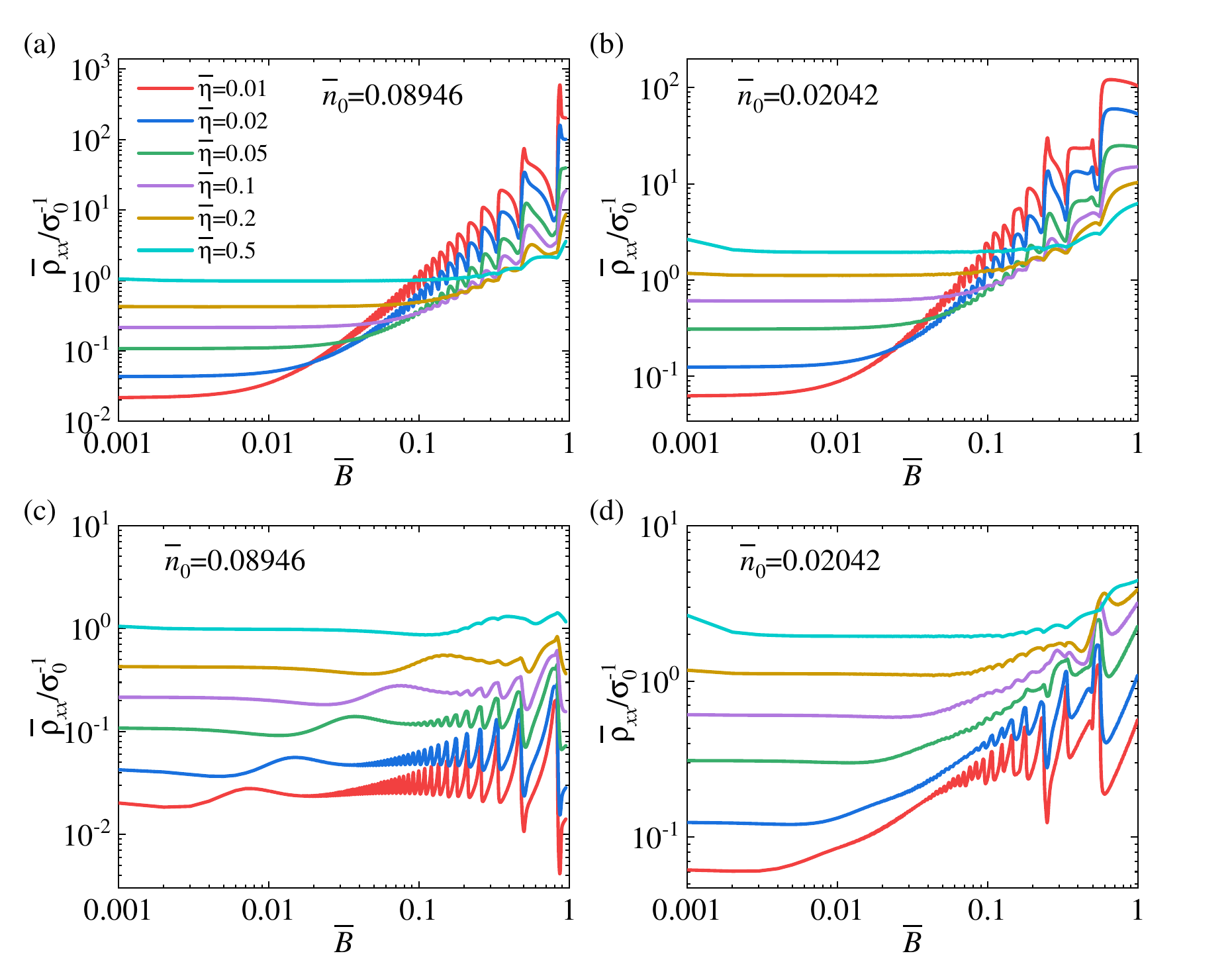}
	\caption{(Color online) The longitudinal MR $\bar\rho_{xx}$ vs the magnetic field $\bar B$.  Note that the double log-axis is used.  The different lines represent the different linewidths $\bar\eta$.  In (a) and (b), $\bar\rho_{xx}$ is calculated as $\bar\rho_{xx}=\bar\sigma_{xx}^{-1}$, while in (c) and (d), $\bar\rho_{xx}$ is calculated by using Eq.~(\ref{rhoxx}).  
	The carrier density is fixed as $\bar n_0=0.02042$ in (a) and (c), and $\bar n_0=0.08946$ in (b) and (d).  The legends are the same in all figures. }
	\label{Fig5}
\end{figure} 

In their theoretical analysis~\cite{S.Lei}, the authors believed that in nodal-line semimetals, the velocity sign flip will result in the negligibly small Hall conductivity, $\sigma_{xy}\simeq0$.  Thus, by neglecting $\sigma_{xy}$, they calculated the MR as $\rho_{xx}=\sigma_{xx}^{-1}$ and obtained the nonsaturating MR with the magnetic field, which seems to be consistent with their experimental results. 
Following this way, we also calculate $\bar\rho_{xx}$ and plot the results in Figs.~\ref{Fig5}(a) and~\ref{Fig5}(b).  
In both figures, we see that at weak $\bar\eta$, $\bar\rho_{xx}$ increases steadily with $\bar B$, indicating the nonsaturating MR behavior in both high-energy and low-energy regions.  
This is understood that with increasing $\bar B$, the LL spacings will get enlarged and less LLs will be involved in the magnetotransport, leading to the nonsaturating $\bar\rho_{xx}$.  

Nevertheless, in Fig.~\ref{Fig4}, since the magnitude of $\bar\sigma_{xy}$ is comparable to $\bar\sigma_{xx}$, $\bar\sigma_{xy}$ may not be simply neglected in calculating $\bar\rho_{xx}$, which is also supported by the calculations within the Boltzmann transport equation (see Sec.I of SM~\cite{SM}).  Based on these analysis, we recalculate $\bar\rho_{xx}$ through Eq.~(\ref{rhoxx}) and plot the results in Figs.~\ref{Fig5}(c) and~\ref{Fig5}(d).  
In Fig.~\ref{Fig5}(c) of the high-energy region, even at weak $\bar\eta$, $\bar\rho_{xx}$ behaves almost unchanged to the magnetic field $\bar B$ until the system enters the quantum oscillation region. 
Then the SdH oscillations of the MR dominate, with the amplitude increasing slowly till $\bar B=0.835$, where the quantum limit is reached that the electrons are confined to the lowest zeroth LLs.  
By contrast, in Fig.~\ref{Fig5}(d) of the low-energy region, at weak $\bar\eta$, $\bar\rho_{xx}$ increases steadily with $\bar B$ and the quantum limit is reached when $\bar B=0.56$.  
For the effect of the impurity scatterings, the increasing $\bar\eta$ will suppress the oscillations.  At $\bar\eta=0.5$, the MR line becomes quite flat with $\bar B$ and there are no oscillations.  

\begin{figure}
	\includegraphics[width=9.2cm]{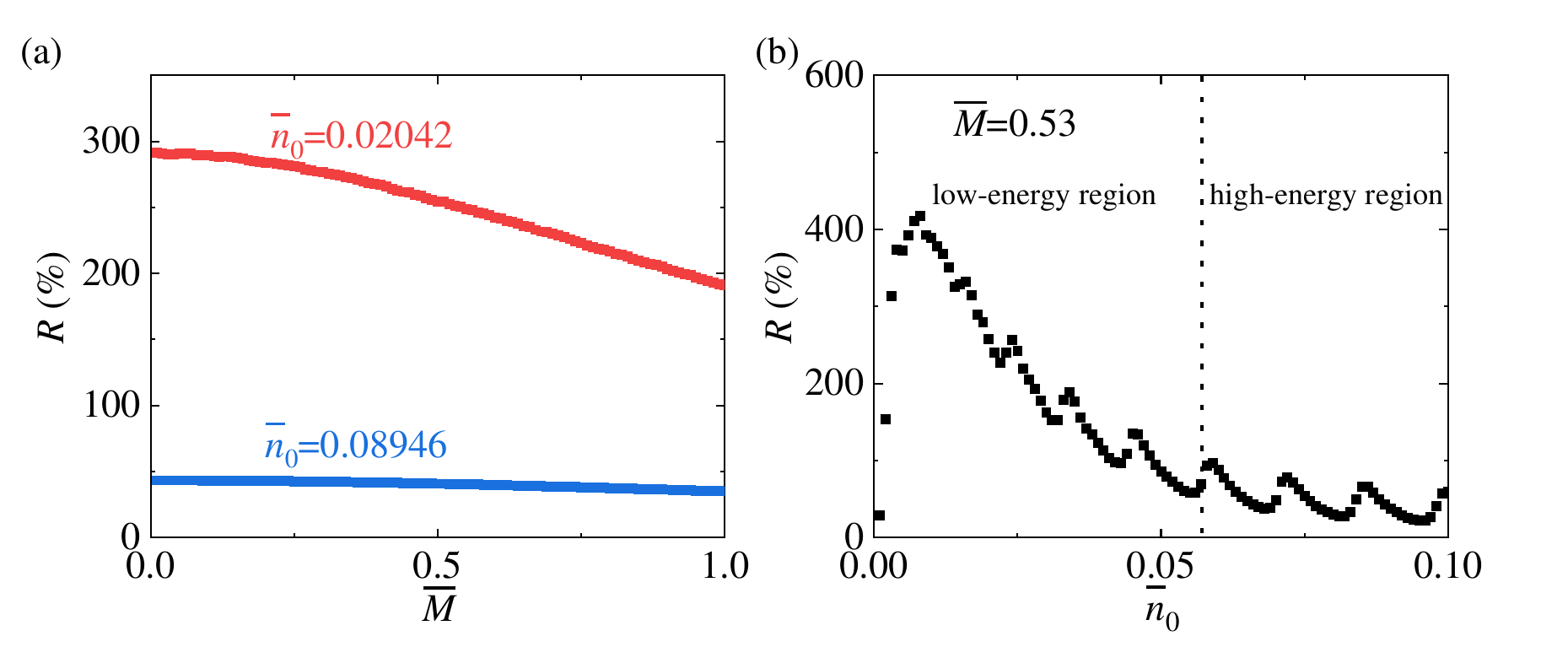}
	\caption{(Color online) The MR ratio $R$ vs the Dirac mass $\bar M$ in (a) and the carrier density $\bar n_0$ in (b).  In (b), the dotted line at $\bar n_0=0.057$ separates the low-energy region from the high-energy region.  The linewidth is chosen as $\bar\eta=0.01$. 	
	In (a), we set $\bar n_0=0.02042$ for the upper red line and $\bar n_0=0.08946$ for the lower blue line; in (b), we set $\bar M=0.53$. } 
	\label{Fig6}
\end{figure} 

To compare with the experiment directly, we calculate the MR ratio $R$ as 
\begin{align}
R=\frac{\bar\rho_{xx}(\bar B_{\text{max}})-\bar\rho_{xx}(\bar B=0)}{\bar\rho_{xx}(\bar B=0)}\times 100\%. 
\label{ratio}
\end{align}
In our theoretical model, when taking the nodal ring radius $k_r=0.1$ \AA$^{-1}$ and the effective mass of the electron $m=0.4m_0$, with $m_0$ the bare electron mass, the energy scale is $\varepsilon_r=0.0944$ eV.  
Then the chosen parameters $\bar M=0.53$ and $\gamma=0.7$ correspond to the Fermi velocity $v_z=10^5$ m/s and the Dirac mass $M=0.05$ eV, respectively.  These model parameters are comparable to those in real nodal-line semimetal material, EuGa$_4$~\cite{S.Lei}.  
The maximum magnetic field that can be reached in the experiment is $B_{\text{max}}=40$ T~\cite{S.Lei}, which corresponds to $\bar B_{\text{max}}=0.061$.  On the other hand, the minimum magnetic field that we use in the calculations is $\bar B_{\text{min}}=0.001$, which corresponds to $B_{\text{min}}=0.655$ T.   
Moreover, as seen in Fig.~\ref{Fig5}, $\bar\rho_{xx}$ keeps unchanged when the magnetic field approaches $\bar B_{\text{min}}$.  We expect that the unchanged value of $\bar\rho_{xx}$ can  persist till $\bar B=0$. 
Thus, in Eq.~(\ref{ratio}), $\bar\rho_{xx}(\bar B=0)$ could be replaced by $\bar\rho_{xx}(\bar B_{\text{min}})$. 
In Fig.~\ref{Fig5}(d), we have $\bar\rho_{xx}(\bar B=0.001)=0.06$ and $\bar\rho_{xx}(\bar B=0.061)=0.2163$ at $\bar\eta=0.01$, from which the MR ratio is obtained as $R\simeq260\%$. 

To further elucidate the magnitude of the MR ratio $R$, we investigate its dependence on the Dirac mass $\bar M$ and the carrier density $\bar n_0$, and the results are plotted in Fig.~\ref{Fig6}.  
Experimentally, the Dirac mass may be controlled in various ways, such as the uniaxial strain, pressure, external electric field, and circularly polarized light~\cite{Z.Yan2016}, while the carrier density varies in different samples~\cite{Y.Okamoto}.  

In Fig.~\ref{Fig6}(a), $\bar M$ is chosen to increase from $\bar M=0$ to $\bar M=1$, which corresponds to the Dirac mass $M=(0\sim0.094)$ eV.  
In the low-energy region $\bar n_0=0.02042$, although $R$ decreases with $\bar M$ from $290\%$ to $200\%$ (see the red line), the nonsaturating MR behavior is still present; while in the high-energy region $\bar n_0=0.08946$, $R$ seems to be less affected by the Dirac mass variation and is about 50\% (see the blue line).  
In Fig.~\ref{Fig6}(b), $\bar n_0$ varies from 0.001 to 0.1, which covers a large range of the carrier density $n_0=(10^{18}\sim10^{20})$ cm$^{-3}$.  
We see that the MR ratio $R$ oscillates with $\bar n_0$, in which the amplitude decreases from $400\%$ to $150\%$ in the low-energy region, and is around $50\%$ in the high-energy region. 
The results indicate that the nonsaturating MR behavior occurs in the low-energy region, but in the the high-energy region, the MR seems to be saturated.  
Compared to the giant value $R\simeq2\times10^5\%$ reported in the experiment~\cite{S.Lei}, the MR ratio obtained in our calculations is much smaller.

\section{Discussions and Summaries}

When extending the nodal-line semimetal model to the nodal-surface semimetal model, the Hamiltonian is written as~\cite{H.K.Pal, L.Zhang, Z.Rukelj}: $\hat H=[\frac{\hbar^2}{2m}(k_x^2+k_y^2+k_z^2)-\varepsilon_r]\sigma_z +M\sigma_x$.
Clearly, the conduction and valence bands will touch on a nodal surface in the BZ when the Dirac mass $M=0$. 	
Our calculations indicate that in the low-energy region, there are no evident oscillation  frequencies of the chemical potential $\bar\mu$.  
This is understood as follows.  When the inverse magnetic field $\bar B^{-1}$ increases, the saddle point at $\bar k_z=0$ of each LL will decrease at first and then increase, which are the same as the torus model. 
This means that each saddle point will cross $\bar\mu$ twice with $\bar B^{-1}$ and there may exist  two oscillation frequencies.  
However, for the $n$-th LL, only the inverted part around $\bar k_z=0$, but not the whole one, will cross $\bar\mu$.  Thus, the scenario is quite different from the picture described by the Onsager’s relation and the torus model.  The results suggest that, even if the two oscillation frequencies may exist, they will be hidden in the background amplitude of the FFT.  More calculation details and analysis can be found in Sec. II of SM~\cite{SM}.  

We discuss the spin Zeeman effect, which will split the bands as the upspin and downspin species.  The corresponding Hamiltonian is written as $\hat H_Z=\frac{1}{2}g\mu_BBs_z$, where $g$ is the Land\'e $g$-factor, $\mu_B$ is the Bohr magneton, and $s_z$ is the Pauli matrice acting on the spin space.  When taking the typical $g$-factor value $g\sim 10$, the splitting energy is estimated as $\Delta\varepsilon=2.9\times10^{-4} B$ eV, which is much smaller than the chosen energy scale $\varepsilon_r$ even when $B=40$ T.  Thus the spin Zeeman splittings are omitted in our magnetotransport calculations.   

To summarize, in this paper, we have studied the magnetotransport property in nodal-line semimetals.  We find that there exist two distinct frequencies in the chemical potential and MC oscillations in the low-energy region, which are attributed to the characteristic torus Fermi surface and can be regarded as an experimental signature of nodal-line semimetals.  
The obtained MR ratio is much smaller than the giant value reported in the experiment~\cite{S.Lei}.  According to these results, we suggest that the giant MR may not arise from the nodal-line semimetal states; instead, it may be caused by some other factors, such as the magnetic property in Eu~\cite{D.Santos-Cottin}.  
The missing orders of magnitude in MR between our theoretical calculations and the experiment could be traced to the missing coupling between the localized Eu magnetic moments and the itinerant carriers, which may be achieved by adding the spin-dependent scatterings into the calculations.  
More magnetotransport studies on nodal-line semimetals are expected in the future.

\section{Acknowledgments} 

This work was supported by the Natural Science Foundation of China (Grant No. 11804122).


\begin{thebibliography}{100}   

\bibitem{B.Q.Lv}
B. Q. Lv, T. Qian, H. Ding, 
Experimental perspective on three-dimensional topological semimetals,  
Rev. Mod. Phys. {\bf93}, 025002 (2021). 

\bibitem{J.A.Sobota}
J. A. Sobota, Y. He, Z. X. Shen, 
Angle-resolved photoemission studies of quantum materials, 
Rev. Mod. Phys. {\bf93}, 025006 (2021).  

\bibitem{A.A.Burkov} 
A. A. Burkov, M. D. Hook, and L. Balents, 
Topological nodal semimetals, 
Phys. Rev. B {\bf84}, 235126 (2011).

\bibitem{M.Phillips} 
M. Phillips and V. Aji, 
Tunable line node semimetals, 
Phys. Rev. B {\bf90}, 115111 (2014). 

\bibitem{C.Fang}
C. Fang, Y. Chen, H. Y. Kee, and L. Fu, 
Topological nodal line semimetals with and without spin-orbital coupling,  
Phys. Rev. B {\bf92}, 081201 (2015). 

\bibitem{Y.H.Chan}
Y. H. Chan, C. K. Chiu, M. Y. Chou, and A. P. Schnyder, 
Ca$_3$P$_2$ and other topological semimetals with line nodes and drumhead surface states, 
Phys. Rev. B {\bf93}, 205132 (2016).  

\bibitem{G.Biana}
G. Bian, T. R. Chang, H. Zheng, S. Velury, S. Y. Xu, T. Neupert, C. K. Chiu, S. M. Huang, D. S. Sanchez, I. Belopolski, N. Alidoust, P. J. Chen, G. Chang, A. Bansil, H. T. Jeng, H. Lin, and M. Z. Hasan, 
Drumhead surface states and topological nodal-line fermions in TlTaSe$_2$, 
Phys. Rev. B {\bf93}, 121113(R) (2016). 

\bibitem{I.Belopolski}
I. Belopolski, K. Manna, D. S. Sanchez, G. Chang, B. Ernst, J. Yin, S. S. Zhang, T. Cochran, N. Shumiya,
H. Zheng, B. Singh, G. Bian, D. Multer, M. Litskevich, X. Zhou, S. M. Huang, B. Wang, T. R. Chang, S. Y. Xu, A. Bansil, C. Felser, H. Lin, M. Z. Hasan, 
Discovery of topological Weyl fermion lines and drumhead surface states in a room temperature magnet, 
Science {\bf365}, 1278 (2019). 

\bibitem{S.T.Ramamurthy}
S. T. Ramamurthy, T. L. Hughes, 
Quasitopological electromagnetic response of line-node semimetals, 
Phys. Rev. B {\bf95}, 075138 (2017). 

\bibitem{S.Barati} 
S. Barati and S. H. Abedinpour, 
Optical conductivity of three and two dimensional topological nodal-line semimetals, 
Phys. Rev. B {\bf96}, 155150 (2017).  
 
\bibitem{C.Wang}
C. Wang, W. H. Xu, C. Y. Zhu, J. N. Chen, Y. L. Zhou, M. X. Deng, H. J. Duan, and R. Q. Wang, 
Anomalous Hall optical conductivity in tilted topological nodal-line semimetals, 
Phys. Rev. B {\bf103}, 165104 (2021).  
 
\bibitem{Y.Huh}
Y. Huh, E. G. Moon, and Y. B. Kim, 
Long-range Coulomb interaction in nodal-ring semimetals, 
Phys. Rev. B {\bf93}, 035138 (2016).  

\bibitem{Y.Shao}
Y. Shao, A. N. Rudenko, J. Hu, Z. Sun, Y. Zhu, S. Moon, A. J. Millis, S. Yuan, A. I. Lichtenstein, D. Smirnov, Z. Q. Mao, M. I. Katsnelson, and D. N. Basov, 
Electronic correlations in nodal-line semimetals, 
Nat. Phys. {\bf16}, 636 (2020). 

\bibitem{Z.Wu}
Z. Wu and Y. Wang, 
Nodal topological superconductivity in nodal-line semimetals, 
Phys. Rev. B {\bf108}, 224503 (2023).

\bibitem{L.S.Xie}
L. S. Xie, L. M. Schoop, E. M. Seibel, Q. D. Gibson, W. Xie, and R. J. Cava, 
A new form of Ca$_3$P$_2$ with a ring of Dirac nodes, 
APL Mater. {\bf3}, 083602 (2015).  

\bibitem{G.Bianb}
G. Bian, T. R. Chang, R. Sankar, S. Y. Xu, H. Zheng, T. Neupert, C. K. Chiu, S. M. Huang, G. Chang, I. Belopolski, D. S. Sanchez, M. Neupane, N. Alidoust, C. Liu, B. Wang, C. C. Lee, H. T. Jeng, C. Zhang, Z.  Yuan, S. Jia, A. Bansil, F. Chou, H. Lin, and M. Z. Hasan, 
Topological nodal-line fermions in spin-orbit metal PbTaSe$_2$, 
Nat. Commun. {\bf7}, 10556 (2016). 

\bibitem{D.Takane}
D. Takane, Z. Wang, S. Souma, K. Nakayama, C. X. Trang, T. Sato, T. Takahashi, and Y. Ando, 
Dirac-node arc in the topological line-node semimetal HfSiS, 
Phys. Rev. B {\bf94} 121108(R) (2016). 

\bibitem{Y.Okamoto}
Y. Okamoto, T. Inohara, A. Yamakage, Y. Yamakawa, and Koshi Takenaka, 
Low carrier density metal realized in candidate line-node Dirac semimetals CaAgP and CaAgAs, 
J. Phys. Soc. Jpn, {\bf85}, 123701 (2016).  

\bibitem{E.Emmanouilidou}
E. Emmanouilidou, B. Shen, X. Deng, T. R. Chang, A. Shi, G. Kotliar, S. Y. Xu, and N. Ni, 
Magnetotransport properties of the single-crystalline nodal-line semimetal candidates CaTX (T=Ag, Cd; X=As, Ge), 
Phys. Rev. B {\bf95}, 245113 (2017). 

\bibitem{Y.H.Kwan}
Y. H. Kwan, P. Reiss, Y. Han, M. Bristow, D. Prabhakaran, D. Graf, A. McCollam, S. A. Parameswaran, and A. I. Coldea, 
Quantum oscillations probe the Fermi surface topology of the nodal-line semimetal CaAgAs, 
Phys. Rev. Res. {\bf2}, 012055(R) (2020). 

\bibitem{H.T.Hirose}
H. T. Hirose, T. Terashima, T. Wada, Y. Matsushita, Y. Okamoto, K. Takenaka, and S. Uji, 
Real spin and pseudospin topologies in the noncentrosymmetric topological nodal-line semimetal CaAgAs, 
Phys. Rev. B {\bf101}, 245104 (2020).   

\bibitem{X.B.Wang}
X. B. Wang, X. M. Ma, E. Emmanouilidou, B. Shen, C. H. Hsu, C. S. Zhou, Y. Zuo, R. R. Song, S. Y. Xu, G. Wang, L. Huang, N. Ni, and C. Liu, 
Topological surface electronic states in candidate nodal-line semimetal CaAgAs, 
Phys. Rev. B {\bf96}, 161112(R) (2017). 

\bibitem{N.Xu}
N. Xu, Y. T. Qian, Q. S. Wu, G. Autès, C. E. Matt, B. Q. Lv, M. Y. Yao, V. N. Strocov, E. Pomjakushina, K. Conder N. C. Plumb, M. Radovic, O. V. Yazyev, T. Qian, H. Ding, J. Mesot, and M. Shi, 
Trivial topological phase of CaAgP and the topological nodal-line transition in CaAg(P$_{1-x}$As$_x$), 
Phys. Rev. B {\bf97}, 161111(R) (2018).

\bibitem{L.M.Schoop}
L. M. Schoop, M. N. Ali, C. Straßer, A. Topp, A. Varykhalov, D. Marchenko, V. Duppel, S. S. P. Parkin, B. V. Lotsch, and C. R. Ast, 
Dirac cone protected by non-symmorphic symmetry and three-dimensional Dirac line node in ZrSiS,
Nat. Commun. {\bf7}, 11696 (2016).

\bibitem{M.Neupane}
M. Neupane, I. Belopolski, M. M. Hosen, D. S. Sanchez, R. Sankar, M. Szlawska, S. Y. Xu, K. Dimitri, N. Dhakal, P. Maldonado, P. M. Oppeneer, D. Kaczorowski, F. Chou, M. Z. Hasan, and T. Durakiewicz, 
Observation of topological nodal fermion semimetal phase in ZrSiS,
Phys. Rev. B {\bf93}, 201104(R) (2016).  

\bibitem{J.Hu}
J. Hu, Z. Tang, J. Liu, X. Liu, Y. Zhu, D. Graf, K. Myhro, S. Tran, C. N. Lau, J. Wei, and Z. Mao, 
Evidence of topological nodal-line fermions in ZrSiSe and ZrSiTe, 
Phys. Rev. Lett. {\bf117}, 016602 (2016).

\bibitem{S.Lei}
S. Lei, K. Allen, J. Huang, J. M. Moya, T. C. Wu, B. Casas, Y. Zhang, J. S. Oh, M. Hashimoto, D. Lu, J. Denlinger, C. Jozwiak, A. Bostwick, E. Rotenberg, L. Balicas, R. Birgeneau, M. S. Foster, M. Yi, Y. Sun and E. Morosan,
Weyl nodal ring states and Landau quantization with very large magnetoresistance in square-net magnet EuGa$_4$, 
Nat. Commun. {\bf14}, 5812 (2023).

\bibitem{M.N.Ali}
M. N. Ali, J. Xiong, S. Flynn, J. Tao, Q. D. Gibson, L. M. Schoop, T. Liang, N. Haldolaarachchige,
M. Hirschberger, N. P. Ong, and R. J. Cava,
Large, non-saturating magnetoresistance in WTe$_2$,
Nature {\bf514}, 205 (2014). 

\bibitem{C.Shekhar}
C. Shekhar1, A. K. Nayak, Y. Sun, M. Schmidt, M. Nicklas, I. Leermakers, U. Zeitler, Y. Skourski, J. Wosnitza, Z. Liu, Y. Chen, W. Schnelle, H. Borrmann, Y. Grin, C. Felser and B. Yan, 
Extremely large magnetoresistance and ultrahigh mobility in the topologicalWeyl semimetal candidate NbP, 
Nat. Phys. {\bf11}, 645 (2015).  

\bibitem{X.Huang}
X. Huang, L. Zhao, Y. Long, P. Wang, D. Chen, Z. Yang, H. Liang, M. Xue, H. Weng, Z. Fang, X. Dai, and G. Chen, 
Observation of the Chiral-Anomaly-Induced Negative Magnetoresistance in 3D Weyl Semimetal TaAs, 
Phys. Rev. X {\bf5}, 031023 (2015). 

\bibitem{B.Fauque}
B. Fauqué, X. Yang, W. Tabis, M. Shen, Z. Zhu, C. Proust, Y. Fuseya, and K. Behnia,
Magnetoresistance of semimetals: The case of antimony, 
Phys. Rev. Mat. {\bf2}, 114201 (2018).  

\bibitem{M.X.Yang}
M. X. Yang, W. Luo, and W. Chen, 
Quantum transport in topological nodal-line semimetals. 
Adv. Phys.: X {\bf7}, 2065216 (2022).

\bibitem{V.Pandey}
V. Pandey, D. Joy, D. Culcer, and P. Bhalla, 
Longitudinal dc conductivity in Dirac nodal line semimetals: Intrinsic and extrinsic contributions, 
Phys. Rev. B {\bf110}, 155108 (2024).

\bibitem{D.Shoenberg}
D. Shoenberg, 
\textit{Magnetic Oscillations in Metals} Cambridge University Press, Cambridge, (1984).

\bibitem{L.Zhang}
L. Zhang, X. Y. Song, and F. Wang, 
Quantum oscillation in narrow-gap topological insulators, 
Phys. Rev. Lett. {\bf116}, 046404 (2016). 

\bibitem{J.Knolle}  
J. Knolle and N. R. Cooper, 
Anomalous de Haas–van Alphen effect in InAs=GaSb quantum wells, 
Phys. Rev. Lett. {\bf118}, 176801 (2017)

\bibitem{Y.X.Wang2023}
Y. X. Wang and Z. Cai, 
Quantum oscillations and three-dimensional quantum Hall effect in ZrTe$_5$, 
Phys. Rev. B {\bf107}, 125203 (2023).

\bibitem{C.Li} 
C. Li, C. M. Wang, B. Wan, X. Wan, H.-Z. Lu, and X. C. Xie,
Rules for phase shifts of quantum oscillations in topological nodal-line semimetals, 
Phys. Rev. Lett. {\bf120}, 146602 (2018). 

\bibitem{L.Oroszlany}
L. Oroszlány, B. Dóra, J. Cserti, and A. Cortijo, 
Topological and trivial magnetic oscillations in nodal loop semimetals, 
Phys. Rev. B {\bf97}, 205107 (2018).

\bibitem{H.Yang}
H. Yang, R. Moessner, and L. K. Lim, 
Quantum oscillations in nodal line systems, 
Phys. Rev. B {\bf97}, 165118 (2018). 

\bibitem{Y.Zhao}
Y. Zhao, H. Liu, C. Zhang, H. Wang, J. Wang, Z. Lin, Y. Xing, H. Lu, J. Liu, Y. Wang, S. M. Brombosz, Z. Xiao, S. Jia, X. C. Xie, and J. Wang, 
Anisotropic Fermi surface and quantum limit transport in high mobility three-dimensional Dirac semimetal Cd$_3$As$_2$, 
Phys. Rev. X {\bf5}, 031037 (2015).  

\bibitem{W.B.Rui}
W. B. Rui, Y. X. Zhao, and A. P. Schnyder, 
Topological transport in Dirac nodal-line semimetals, 
Phys. Rev. B {\bf97}, 161113(R) (2018).

\bibitem{T.Devakul}
T. Devakul, Y. H. Kwan, S. L. Sondhi, S. A. Parameswaran, 
Quantum oscillations in the zeroth Landau level: serpentine Landau fan and the chiral anomaly, 
Phys. Rev. Lett. {\bf127} 116602 (2021).

\bibitem{Y.X.Wang2022}  
Y. X. Wang and F. Li, 
Unconventional optical selection rules in ZrTe$_5$ under an in-plane magnetic field, 
Phys. Rev. B {\bf106} 205102 (2022). 

\bibitem{A.Bastin}
A. Bastin, C. Lewiner, O. Betbeder-Matibet, and P. Nozieres, 
Quantum oscillations of the Hall effect of a fermion gas with random impurity scattering, 
J. Phys. Chem. Solids. {\bf32}, 1811 (1971).

\bibitem{L.Smrcka}
L. Smrcka and P. Streda, 
Transport coefficients in strong magnetic fields, 
J. Phys. C: Solid State Phys. {\bf10}, 2153 (1977). 

\bibitem{Z.Cai}
Z. Cai and Y. X. Wang, 
Magnetic field driven Lifshitz transition and one-dimensional Weyl nodes in three-dimensional pentatellurides, 
Phys. Rev. B {\bf108} 155202 (2023).

\bibitem{Abrikosov} 
A. A. Abrikosov, 
Quantum magnetoresistance, 
Phys. Rev. B {\bf58}, 2788 (1998).

\bibitem{V.Konye}
V. K\"onye and M. Ogata, 
Magnetoresistance of a three-dimensional Dirac gas, 
Phys. Rev. B {\bf98}, 195420 (2018).

\bibitem{H.W.Wang}
H. W. Wang, B. Fu, and S. Q. Shen, 
Theory of the anomalous Hall effect in the transition metal pentatellurides ZrTe$_5$ and HfTe$_5$, 
Phys. Rev. B {\bf108}, 045141 (2023). 
	
\bibitem{SM}
See Supplemental Materials for more calculation details and analysis, including: (i) the conductivity within the Boltzmann transport equation, (ii) the oscillation frequency in 3D band inversion model. 

\bibitem{Z.Yan2016}
Z. Yan and Z. Wang, 
Tunable Weyl points in periodically driven nodal line semimetals, 
Phys. Rev. Lett. {\bf117}, 087402 (2016). 

\bibitem{H.K.Pal}
H. K. Pal, F. Pi\'echon, J. N. Fuchs, M. Goerbig, and G. Montambaux, 
Chemical potential asymmetry and quantum oscillations in insulators, 
Phys. Rev. B {\bf94}, 125140 (2016). 	
	
\bibitem{Z.Rukelj}
Z. Rukelj and A. Akrap, 
Carrier concentrations and optical conductivity of a band-inverted semimetal in two and three dimensions, 
Phys. Rev. B {\bf104}, 075108 (2021). 

\bibitem{D.Santos-Cottin}
D. Santos-Cottin, I. Mohelsk\'y, J. Wyzula, F. Le Mardel\'e, I. Kapon, S. Nasrallah, N. Bariši\'c, I. Živkovi\'c, J. R. Soh, F. Guo, K. Rigaux, M. Puppin, J. H. Dil, B. Gudac, Z. Rukelj, M. Novak, A. B. Kuzmenko, C. C. Homes, Tomasz Dietl, M. Orlita, and A. Akrap, 
EuCd$_2$As$_2$: A magnetic semiconductor, 
Phys. Rev. Lett. {\bf131}, 186704 (2023).  


\end{thebibliography}
\end{document}